\documentclass[twoside,leqno,11pt]{article}

\usepackage{proc2e}
\usepackage{alltt}
\usepackage{code}

\newcommand{\email}[1]{{\sf #1}}
\newenvironment{CE}{%
\begin{small}\begin{alltt}}{%
\end{alltt}\end{small}}

\newcommand{\C}[1]{\begin{tt}#1\end{tt}}

\newcommand{\refsec}[1]{Section~\ref{#1}}
\newcommand{\cut}[1]{}
\newcommand{\semantics}[1]{[\hspace{-0.17em}[ #1 ]\hspace{-0.17em}]}

\begin{document}
\cleardoublepage

\title{C++ Templates as Partial Evaluation}

\author{Todd L. Veldhuizen\thanks{Indiana University Computer Science Department, Bloomington Indiana 47405, USA.  Phone: (812) 855-8305.
\email{tveldhui@extreme.indiana.edu}}}

\date{}
\maketitle

\pagenumbering{arabic}

\begin{abstract}
\noindent
This paper explores the relationship between C++ templates
and partial evaluation.
Templates were designed to support generic
programming, but unintentionally provided
the ability to perform compile-time computations and
code generation.  These features are 
completely accidental, and as a result their syntax
is awkward.
By recasting these features in terms of partial evaluation,
a much simpler syntax can be achieved.
C++ may be regarded
as a two-level language in which types are first-class
values.  Template instantiation resembles an offline
partial evaluator.
This paper describes preliminary work toward a single
mechanism based on Partial Evaluation which unifies
generic programming, compile-time computation
and code generation.  The language
Catat is introduced to illustrate these ideas.
\end{abstract}

\section{Introduction}

Templates were added to the C++ language to support
generic programming.  However, their addition
unintentionally introduced powerful mechanisms
for compile-time computation and code generation.
These mechanisms have proven themselves very
useful in generating optimized code for scientific
computing applications 
\cite{POOMA,mtl_iscope,Veldhuizen98a,Veldhuizen:1996:LAC}.  
Since they are accidental
features, their syntax is somewhat awkward.
The goal of this paper is to achieve a simpler
syntax by recasting these features as partial
evaluation.  
We start by briefly summarizing the capabilities provided
by C++ templates, both intended and accidental.



\label{s:capabilities}

\subsection{Generic programming}

The original intent of templates was to support
generic programming, which can be summarized as
``reuse through parameterization''.  Generic
functions and objects have parameters which
customize their behavior.  These parameters must
be known at compile time (i.e. have static binding).
For example, a generic vector class can be declared 
as:

\begin{verbcode}
    template<typename T, int N>             
    class Vector \{                         
       // some member functions here...

    private:
        T data[N];                         
    \};

    // Example use of Vector
    Vector<int,4> x;                       
\end{verbcode}

\noindent
The \C{Vector} class takes two template parameters
(line 1):
\C{T}, a {\it type parameter}, specifies the
element type for the vector; \C{N}, a {\it nontype parameter},
is the length of the vector.
To use the vector class, template arguments must
be provided (line 10).  This causes the template
to be {\it instantiated}: an instance of the
template is created by replacing
all occurrences of \C{T} and
\C{N} in the definition of \C{Vector} 
with \C{int} and \C{4}, respectively.

Functions may also be templates.  Here is a function
template which sums the elements of an array:

\begin{CE}
    template<typename T>
    T sum(T* array, int numElements)
    \{
        T result = 0;
        for (int i=0; i < numElements; ++i)
            result += array[i];
        return result;
    \}
\end{CE}

\noindent
This function works for built-in types, such as
\C{int} and \C{float}, and also for user-defined
types provided they
have appropriate operators (\C{=}, \C{+=}) defined.
Templates allow programmers to develop classes
and functions which are very customizable, yet
retain the efficiency of statically configured
code.  

\subsection{Compile-time computations}

Templates can be exploited to
perform computations at compile time.  This was 
discovered by Erwin Unruh \cite{Unruh94}, who wrote
a program which produced these errors at compile time:

\begin{CE}
    erwin.cpp 10: Cannot convert 'enum' to 'D<2>' 
    erwin.cpp 10: Cannot convert 'enum' to 'D<3>' 
    erwin.cpp 10: Cannot convert 'enum' to 'D<5>' 
    erwin.cpp 10: Cannot convert 'enum' to 'D<7>' 
    erwin.cpp 10: Cannot convert 'enum' to 'D<11>' 
    ...
\end{CE}

\noindent
The program tricked the compiler into calculating a list
of prime numbers!  This capability was quite accidental,
but has turned out to be very useful.  Here is a simpler
example which calculates \C{pow(X,Y)} at compile time:

\label{ctimepow}

\begin{CE}
    template<int X, int Y>
    struct ctime\_pow \{
      static const int result = X * ctime\_pow<X,Y-1>::result;
    \};

    // Base case to terminate recursion
    template<int X>
    struct ctime\_pow<X,1> \{
      static const int result = X;
    \};

    // Example use:
    const int z = ctime\_pow<5,3>::result;  // z = 125
\end{CE}

\noindent
The first template defines a structure \C{ctime\_pow} which
has a single data member \C{result}.  The \C{static const}
qualifiers of \C{result} make its value available
at compile time.  \C{ctime\_pow<X,Y>} refers to
\C{ctime\_pow<X,Y-1>}, so the compiler must recursively
instantiate the template for \C{Y},\C{Y-1}, \C{Y-2}, ...
until it hits the base case provided by the second
template, which is a {\it partial specialization}.

Here is an array class which uses \C{ctime\_pow} to
calculate the number of array elements needed:

\begin{CE}
    template<typename T\_numtype, int N\_length, int N\_dim>
    class SquareArray \{
       // ...
       static const int numElements = ctime\_pow<N\_length,N\_dim>::result;
       T_numtype data[numElements];
    \}

    // Example use:
    SquareArray<float,4,2> x;   // A 4x4 array: will have 16 elements
    SquareArray<float,4,3> x;   // A 4x4x4 array: will have 64 elements
\end{CE}

\noindent
When the \C{SquareArray} template is instantiated, \C{ctime\_pow}
is used to calculate the array size required.
Similar techniques can be used to find greatest
common divisors, test for primality, and so on.  
It is even possible to implement an interpreter for
a subset of Lisp which runs at compile time \cite{CE98}.

\subsection{Code generation}

It turns out that
compile-time versions of flow control structures (loops, 
if/else, case switches) can all be implemented in terms of 
templates.  For example, the definition
of \C{ctime\_pow} (\refsec{ctimepow}) emulates a \C{for} loop using
tail recursion. 
These compile-time programs can perform code generation
by selectively inlining code as they are ``interpreted''
by the compiler.  This technique is called {\it template
metaprogramming} \cite{Veldhuizen95a}.  Here is a 
template metaprogram which generates a specialized
dot product algorithm:

\label{metadot}

\begin{CE}
    template<typename T, int I, int N>
    struct meta\_dot \{
        static inline T f(T* a, T* b)
        \{ return meta\_dot<T,I-1,N>::f(a,b) + a[I]*b[I]; \}
    \};

    template<class T, int N>
    struct meta\_dot<T,0,N> \{
        static inline T f(T* a, T* b)
        \{ return a[0]*b[0]; \}
    \};

    // Example use:
    float x[3], y[3];
    float z = meta\_dot<float,2,3>::f(x,y);   // **
\end{CE}

\noindent
In the above example, the call to \C{meta\_dot} in
line marked \C{**} results in code equivalent to:

\begin{CE}
    float z = a[0]*b[0] + a[1]*b[1] + a[2]*b[2];
\end{CE}

\noindent
Head recursion is used to unroll the loop over the
vector elements.
The syntax for writing such code generators
is clumsy.  However, the technique has proven very useful
in producing specialized algorithms for scientific
computing \cite{mtl_iscope,Veldhuizen:1996:LAC}.

It is even possible to create and manipulate static data
structures at compile time, by encoding them as templates.
This is the basis of the {\it expression templates}
technique \cite{Veldhuizen95b}, which creates parse
trees of array expressions at compile time.  These
parse trees are used to generate efficient evaluation
routines for array expressions.  This technique is
the backbone of several libraries for
object-oriented numerics \cite{POOMA,Veldhuizen98a}.


\subsection{Traits}

The {\it traits} technique \cite{Myers95} allows
programmers to define ``functions'' which operate
on and return {\it types} rather than data.
As a motivating example, consider a generic function
which calculates the average value of an
array.  What should its return type be?
If the array contains integers, a floating-point
result should be returned.  But a floating-point
return type obviously will not suffice for a complex-valued
array.

The solution is to define a traits class which
maps from the type of the array elements to a
type suitable for containing their average.
Here is a simple implementation:

\label{average}
\label{averagetraits}

\begin{CE}
    template<typename T>
    struct average\_traits \{
      typedef T T\_average;      // default behaviour: T -> T
    \};

    template<>
    struct average\_traits<int> \{
      typedef float T\_average;  // specialization:  int -> float
    \};
\end{CE}

\noindent
An appropriate type for averaging an array of
type \C{T} is given by \C{average\_traits<T>::T\_average}.
Here is an implementation of \C{average}:

\begin{CE}
    template<class T>
    typename average\_traits<T>::T\_average average(T* array, int N)
    \{
        typename average\_traits<T>::T\_average result = sum(array,N);
        return result / N;
    \}
\end{CE}

\section{Templates as partial evaluation}

\label{s:templates-as-PE}

Partial evaluators \cite{Jones:1996:IPE} 
regard a program's
data as containing two subsets: static data, which is known at
compile time, and dynamic data, which is not known until
run time.  A partial evaluator evaluates as much of a
program as possible (using the static data) and outputs
a specialized {\it residual} program.

To determine which portions of a program may be
evaluated, a partial evaluator performs
{\it binding time analysis} to label language
constructs and data as static or dynamic.
Such a labelled language is called
a {\it two-level language}.
For example, a binding-time analysis of some
scientific computing code might produce this two-level
code fragment:

\begin{CE}
    float volumeOfCube(float length)
    \{
        return pow(length,\underline{3});
    \}

    float pow(float x, \underline{int N})
    \{
        float y = 1;
        \underline{for} (\underline{int i=0}; \underline{i < N}; \underline{++i})
            y *= x;
        return y;
    \}
\end{CE}

\noindent
in which static constructs have been underlined.
A partial evaluator such as CMix \cite{Andersen:94:PhD} would 
evaluate the static constructs to produce the residual
code:

\begin{CE}
    float volumeOfCube(float length)
    \{
        return pow3(length);
    \}

    float pow3(float x)
    \{
        float y = 1;
        y *= x;
        y *= x;
        y *= x;
    \}
\end{CE}

\noindent
Such specializations can result in substantial performance
improvements for scientific code \cite{Berlin90,GlueckNakashigeZoechling:95:IFIP}.

\subsection{C++ as a two-level language}

\noindent
C++ templates resemble a two-level language.
Function templates take both template parameters
(which have static binding) and function arguments
(which have dynamic binding).  For example,
the \C{pow} function of the previous example
might be declared in C++ as:

\begin{CE}
    template<int N>
    float pow(float x);     // Calculate pow(x,N)
\end{CE}

\noindent
The static data (\C{N}) is a template parameter, and the
dynamic data (\C{x}) is a function argument.  
To incorporate template type parameters into this viewpoint, we need 
to regard types as first-class values.  For example, in
a declaration such as

\begin{CE}
    template<typename X, int Y>
    void func(int i, int j);
\end{CE}

\noindent
we regard \C{X} as a piece of data whose value is a type.
Since C++ is statically typed, type variables may only
have static binding.

It is useful to think of the type of \C{X} as being
\C{typename}, which can be regarded as a type whose
possible values span all types.
This point of view has a certain simplifying
power: for example, we can now view
\C{typedef}s as assignments
between typename variables.  For example,

\begin{CE}
    typedef float float\_type;
\end{CE}

\noindent
can be regarded as equivalent to the (fictional syntax)

\begin{CE}
    typename float\_type = float;
\end{CE}

\subsection{Template instantiation as offline PE}

Partial evaluation of languages which contain binding-time
information is called {\it offline} partial evaluation.
Template instantiation resembles offline
partial evaluation: 
the compiler takes template code (a two-level language)
and evaluates those portions of the template which 
involve template parameters (statically bound values).
For example, consider this template class:

\begin{CE}
template<int X>
struct foo \{
    static const int result = foo<(X \% 2 == 0) ? (X/2) : (3*X+1)>::result;
\};

// Base case: X = 1
template<>
struct foo<1> \{
    static const int result = 0;
\};
\end{CE}

\noindent
When \C{foo<X>} is instantiated, the compiler must determine
if \C{X \% 2 == 0} (i.e. whether \C{X} is even). If true,
it instantiates \C{foo<X/2>}; otherwise \C{foo<3*X+1>} is
instantiated.  In theory, this continues until the compiler hits
the base case \C{X=1}.\footnote{Whether this recursion terminates
for all \C{X} is a well-known open problem.  In C++, it
is impossible to determine if a chain of template instantiations
will ever terminate.  For this reason, compilers place limits
on the depth of template instantiation chains.}


\section{A simpler syntax: Catat}

\label{s:Catat}

We now present preliminary ideas for a single
mechanism based on Partial Evaluation 
which unifies generic programming,
compile-time computation, and code generation.
To illustrate the ideas, we introduce a
(currently hypothetical) language
Catat.  Catat is a multi-level language based on C++ in which
types are first-class values.  

\subsection{Binding time specifications}

Each scope in a Catat program is associated with
a default binding time.  By default, the global
scope has dynamic binding.  To indicate statically
bound variables, an \C{@} symbol is appended to
the type:

\begin{CE}
    int i = 0;   // Dynamic data
    int@ j = 0;  // Static data
\end{CE}

\noindent
The type \C{int@} is equivalent to \C{const}
\C{int} in C++.  

To preserve consistency between the dynamic and static
versions of the language, it is necessary to allow
multiple levels of binding (or {\it stages}).  The
\C{@} symbol indicates that a variable is bound in the
previous stage.

The \C{@} symbol may also be applied to control
constructs:

\begin{CE}
    // Calculate N! (factorial) at compile time
    int@ N = 5, Nfact = 1;
    for@ (int@ i=1; i < N; ++i)
        Nfact *= i;
\end{CE}

\noindent
Operators such as \C{=} and \C{*} are 
applied at compile-time if their operands are statically
bound.
Data may flow from static to dynamic
constructs, but not vice-versa.  This is called
cross-stage persistence by Taha and Sheard
\cite{TahaSheard:97:multi-stage}.
For example:

\begin{CE}
    int@ i;
    int j;

    j = i;   // Okay, i is known at runtime
    i = j;   // Not okay, j not known at ctime
\end{CE}

\subsection{Functions}

\noindent
Functions in Catat may take a mixture of static
and dynamic arguments.  We find it convenient to
give functions two separate parameters lists, as in C++.
Here is an implementation of the \C{meta\_dot}
function described earlier:

\begin{CE}
    function dot(int@ N, typename@ T)(T* a, T* b) \{
        T result = 0;
        for@ (int@ i=0; i < N; ++i)
            result += a[i]*b[i];
        return result;
    \}
\end{CE}

\noindent
Note how much simpler this definition is than its
template metaprogram counterpart (\refsec{metadot}).

The function \C{dot} may be thought of as a
{\it generating extension} or higher-order
function.  The concept is easier to express
in a functional notation:

\begin{CE}
    (define dot
      (lambda (static-parms)
        (PE static-parms
          (lambda (dynamic-parms)
            body))))
\end{CE}

\noindent
where \C{(PE parms expr)} performs partial evaluation of
\C{expr} using static parameters \C{parms}.  The use of
argument lists of the form \C{(static-parms)(dynamic-parms)}
hints at this idea, and also avoids the parsing difficulty
associated with \C{<>} brackets in C++.

Catat discards the return type specification of C++
and replaces it with the keyword \C{function}.  The return
type may result from compile-time calculations, and
so must be inferred from the body of the function.
As in C++, we allow static parameters to be inferred
from dynamic argument types; for example, in the
function \C{average}, \C{T} can be inferred from the
type of the \C{array} argument.

Functions may be evaluated at either compile-time
or run-time.  They are not fixed to any stage.  
For example, given this definition of \C{pow}:

\begin{CE}
    function pow(int X, int N) \{
        int result = 1;
        for (int i = 0; i < N; ++i)
            result *= X;
        return result;
    \}
\end{CE}

\noindent
One can invoke \C{pow} at both run-time and compile-time:

\begin{CE}
    int result1 = pow(2,3);   // Evaluated at run-time
    int@ result2 = pow@(2,3); // Evaluated at compile-time
\end{CE}

\noindent
Functions can replace the use of traits classes in C++.
Here is a Catat version
of \C{average\_traits} (\refsec{averagetraits}):

\begin{CE}
    // Return a type appropriate for averaging an array of T
    function average\_type(typename T) \{
        switch(T) \{
            case int:       return float;
            case char:      return float;
            case long int:  return double;
            // etc.
            
            default:        return T;
        \}
    \}
\end{CE}

\noindent
This illustrates the usefulness of regarding types as
first-class values.
Here is the \C{average} function, recoded in Catat:

\begin{CE}
    function average(typename@ T)(T* array, int N) \{
        typename@ T_average = average\_type@(T);
 
        // Sum the array, divide by N 
        T_average sum = 0;
        for (int i=0; i < N; ++i)
            sum += array[i];
        return sum / N;
    \}
\end{CE}

\subsection{Specialization}

When calls to function templates are encountered
during C++ compilation, the template is
instantiated.  In Catat a similar process
would occur, which may be called
{\it specialization}: a partial evaluator
produces a residual function by evaluating
the static constructs.  This function call:

\begin{CE}
    int data[10];   // ..
    float result = average(int)(data,10);
\end{CE}

\noindent
triggers the partial evaluation of \C{average}; the
resulting specialization (translated into C++) might be

\begin{CE}
    float average__int(int* array, int N) \{
        float sum = 0;
        for (int i=0; i < N; ++i)
            sum += array[i];
        return sum;
    \}
\end{CE}

\subsection{Classes}

Classes in Catat may take static parameters, and contain both
static and dynamic data members.
For example:

\codelinereset
\begin{verbcode}
    class SquareArray(typename@ T\_numtype, int@ N\_length, int@ N\_dim) \{

    public:
        SquareArray@()
        \{
            // Calculate array size needed
            if@ ((N\_dim < 1) || (N\_length < 1))
                Catat\_error@("N\_dim and N\_length must be positive.");
            else@
                numElements = pow@(N\_length,N\_dim);
        \}

        SquareArray()
        \{
            // Initially set elements to zero
            for (int i=0; i < numElements; ++i)
                data[i] = 0;
        \}

    private:
        static int@ numElements;
        T_numtype data[numElements];
    \}
\end{verbcode}

\noindent
In this class, there are two constructors: a compile-time
constructor \C{SquareArray@()} and a runtime constructor
\C{SquareArray()}.  The \C{SquareArray@()} constructor is
invoked during compilation when the class is
specialized.  At this time, it can check the static
parameters to ensure they are correct; if not, a
compile-time error is issued.  With the aid of some
reflection, this may be the right way to
enforce constraints on template parameters (an idea due
to Vandevoorde \cite{Czarnecki98b}).
The constructor
\C{SquareArray()} is invoked at runtime when instances
of the class are created.

As with functions, classes have no specified binding time,
but may be instantiated at any stage.  For example,

\begin{CE}
SquareArray@(int,3,2) x;  // Array instantiated at compile-time
SquareArray(int,3,2) y;   // Array instantiated at run-time
\end{CE}

\subsection{Thoughts on compiling Catat}

To implement Catat as described, one apparently needs both
a Catat-interpreter and a Catat-compiler.  For example, to
compile a function $f$ which has static parameters $s$
and dynamic parameters $d$, these steps would be needed:

\noindent
1. Use the interpreter to partially evaluate $f$ using
the static parameters $s$:

\begin{eqnarray}
\semantics{interp} [f,s] = f_s \nonumber
\end{eqnarray}

\noindent
2. Use the compiler to produce native code for
the residual function $f_s$:

\begin{eqnarray}
\semantics{compiler} f_s = f_s^\star \nonumber
\end{eqnarray}

\noindent
where $\star$ indicates native code.\footnote{The $\semantics{\cdot}$
notation is from partial evaluation: if $pow$ is a function,
then $\semantics{pow}[input]$ is the result of executing
$pow$ with $input$.  For example, $\semantics{pow}[3,2]=9$.}
This seems wasteful;
there would be lots of duplicated effort to create both
an interpreter and compiler.

It may be possible to avoid this problem by using an approach
similar to that pioneered by the Cmix partial evaluation system
\cite{Andersen:94:PhD}.  The basic approach is to use
a ``closure compiler'' which uses run-time code generation (RTCG)
to compile a single function.  RTCG is a bit of a misnomer, since
the code generation is being done at compile-time by the compiler.

For example, to evaluate code such as

\begin{CE}
    int@ result2 = pow@(2,3); // Evaluated at compile-time
\end{CE}

\noindent
One could use the closure compiler (CC) to compile \C{pow}
to native code, and then execute the function with the
arguments $(2,3)$:

\begin{eqnarray}
\semantics{CC}pow & = & pow^\star \nonumber \\
\semantics{pow^\star}[2,3] & = & 8 \nonumber
\end{eqnarray}

\noindent
To evaluate a two-level function such as \C{f(s)(d)}, it
must first be transformed into a single-level function.
We suggest the term {\it flattening} for this transform.
Flattening turns two-level code into single-level
code, 
by replacing dynamic code with static code that generates
syntax trees for the dynamic code.  For example, the
two-level function

\begin{CE}
    float pow(int@ N)(float x)
    \{
        float result = 1;
        for@ (int@ i = 0; i < N; ++i)
            result *= x;
        return result;
    \}
\end{CE}

\noindent
would be transformed into something like:

\begin{CE}
    ASTree powgen(int N)
    \{
        ASTree func = make_lambda(...);
        ASTree x = make_varref("x");

        func.body().append(make_vardecl(float, "result", 1));
        ASTree result = make_varref("result");

        for (int i=0; i < N; ++i)
            func.body().append(make_op("*=", result, x));

        func.body().append(make_return(result));
        return fn;
    \}
\end{CE}

\noindent
With the aid of this flattening transformation, two-level
functions can be compiled without an interpreter:

\begin{eqnarray}
f(s)(d) & \overrightarrow{{\tiny flatten}} & f_{gen}(s) \nonumber \\
\semantics{CC} f_{gen} & = & f_{gen}^\star  \nonumber \\
\semantics{f_{gen}^\star} s & = & f_s \nonumber \\
\semantics{CC} f_s & = & f_s^\star \nonumber
\end{eqnarray}

\noindent
i.e. first the flattening transform is used to turn $f(s)(d)$
into a single-level generator for $f$, called $f_{gen}$.  This generator
is compiled using $CC$ into native code $f_{gen}^\star$.  The native-code
version is then executed with the static parameters $s$, and produces
the specialized version $f_s$.  This function is then compiled using
$CC$.

The availability of fast, portable run-time code generation
systems such as \C{vcode} \cite{pldi96*160} makes this approach
to compilation possible.

\subsection{Interesting possibilities}

Languages with Catat-like abilities raise some interesting possibilities:


\subsubsection{Scripting}

The partial evaluator for Catat needs to contain
what is essentially an interpreter to evaluate the
static portions of the program.  This implies that
you get scripting for no extra cost; a Catat
program consisting solely of static constructs
will be completely interpreted, with no residual
code generated.  With a little bit of extra
work, it ought to be possible to dynamically
link to already-compiled Catat programs;
this would make it possible to ``steer''
applications using a natural scripting interface.

\subsubsection{Futamura projection}

Suppose we wrote an interpreter for a domain-specific
language (DSL) in Catat.
We could design our interpreter to take the input
text as a static parameter, and input
variables as dynamic parameters.  Residualization
of the interpreter would result in the DSL
code being compiled into the dynamic subset
of Catat, via the first Futamura projection \cite{Futamura:71}.
This approach would allow users to incorporate
fragments of domain-specific languages into
their applications, without sacrificing
efficiency.

\subsubsection{Reflection and Meta-level Processing}

A language like Catat may provide a natural environment
for implementing reflection and meta-level processing
capabilities, since the ability to perform compile-time
calculations is there already.  Such capabilities would
allow programmers to query objects about their methods
and members, determine the parameter types of
functions, and perhaps even manipulate and generate abstract
syntax trees.






\section{Related Work}

Nielson and Nielson~\cite{Nielson-Nielson92a}
first investigated two-level languages
and showed that binding-time analysis can be expressed
as a form of type checking.  
The most closely related work is MetaML, a statically typed 
multi-level language for hand-writing
code generators \cite{TahaSheard:97:multi-stage}.
MetaML does not appear to address the issue of generic
programming.
Gl\"{u}ck and J{\o}rgensen described a program generator for multi-level
specialization~\cite{GlueckJoergensen:97:multi-spec}
which uses a multi-level functional language 
to represent automatically produced program generators.

Metalevel processing systems address many of the same
problems as Catat; they give library writers the ability
to directly manipulate abstract syntax trees at compile time.  
Relevant examples are Xroma \cite{Czarnecki98b}, MPC++ \cite{Ishikawa96},
Open C++ \cite{OpenC++}, and Magik \cite{EnglerDSL}.  
These systems are not phrased in terms of partial evaluation or
two-level languages;
code generation is generally done by constructing abstract
syntax trees.  A more closely related system is Catacomb
\cite{Stichnoth97}, which provides a two-level language for
generating runtime library code for parallelizing compilers.  However, it
does not address issues of generic programming.

The idea of types as first-class values originates in 
polymorphic or second order typed lambda calculus,
and languages based on it. 

\section{Conclusions}

We have shown that C++ with templates may be regarded as a
two-level language in which types are first-class values
with static binding, and that template instantiation 
bears a striking resemblance to offline partial evaluation.
Languages built on this insight may offer a way
to provide generic programming, code generation, and
compile-time computation via a single mechanism
with simple syntax.

\section{Acknowledgments}

This work was supported in part by NSF grants CDA-9601632
and CCR-9527130.  We are grateful to Robert Gl{\"u}ck for
pointing out relevant work in Partial Evaluation.



\bibliography{robert,partial,ulrich,oon,activelib,rtcg}
\bibliographystyle{siamproc}  

\end{document}